\documentclass[10pt, twocolumn]{IEEEtran}
\usepackage{graphicx}
\usepackage{epsfig}
\usepackage{algorithm}
\usepackage{algorithmic}
\usepackage{ifmtarg}
\usepackage{cite}
\usepackage{amssymb}
\usepackage{amsthm}
\usepackage[fleqn]{amsmath}
\usepackage{amsmath}
\usepackage{color}
\usepackage{bbm}
\usepackage{cite}
\usepackage{epstopdf}
\usepackage[table,xcdraw]{xcolor}
\usepackage{flushend}
\newtheorem{theorem}{\textbf{\text{Theorem}}}
\newtheorem{lemma}{\textbf{\text{Lemma}}}

\newtheorem{corollary}{Corollary}

\usepackage{flushend}
\usepackage{subcaption}

\begin{document}
\title{Modeling Cellular Networks in Fading Environments with Dominant Specular Components}
\author{ Ahmad AlAmmouri, Hesham ElSawy, Ahmed Sultan-Salem, Marco Di Renzo, and Mohamed-Slim Alouini\\ 
\thanks{ Ahmad AlAmmouri, Hesham ElSawy, Ahmed Sultan-Salem and Mohamed-Slim Alouini are with
King Abdullah University of Science \& Technology (KAUST),  Saudi Arabia. Emails: \{ahmad.alammouri, hesham.elsawy, ahmed.salem, slim.alouini\}@kaust.edu.sa} 
\thanks{  Marco Di Renzo is with Paris-Saclay University - Signals \& Systems Lab. (CNRS - CentraleSupelec - Univ Paris-Sud) - France. Email: marco.direnzo@lss.supelec.fr}
}

\maketitle

\begin{abstract}
Stochastic geometry (SG) has been widely accepted as a fundamental tool for modeling and analyzing cellular networks. However, the fading models used with SG analysis are mainly confined to the simplistic Rayleigh fading, which is extended to the Nakagami-m fading in some special cases. However, neither the Rayleigh nor the Nakagami-m accounts for dominant specular components (DSCs) which may appear in realistic fading channels. In this paper, we present a tractable model for cellular networks with generalized two-ray (GTR) fading channel. The GTR fading explicitly accounts for two DSCs in addition to the diffuse components and offers high flexibility to capture diverse fading channels that appear in realistic outdoor/indoor wireless communication scenarios. It also encompasses the famous Rayleigh and Rician fading as special cases. To this end, the prominent effect of DSCs is highlighted in terms of average spectral efficiency.
\end{abstract}

\section{Introduction}
Modern cellular networks have evolved from the ubiquitous hexagonal grid to irregular multi-tier structure that randomly changes from one geographical location to another \cite{tractable_app, martin_ppp, marco_fitting}. To cope with such evolution, efforts are spent to develop tractable stochastic geometry (SG) models that account for large-scale spatial randomness as well as different sources of uncertainties that emerge in modern cellular networks, such as multipath fading, shadowing, and power control~\cite{survey_h}.  The last decade witnessed significant progress developing all aspects of the SG models, except for the fading environments. In the context of multi-path fading, the Rayleigh fading assumption is almost a common factor in the literature. In addition to the Rayleigh fading case, there are some proposals that incorporate Nakagami-m fading into tractable SG analysis~\cite{eid_Mimo, Laila_letter}.

Both the Rayleigh and Nakagami-m assumptions are favorable because they lead to a desirable exponential expressions for the conditional (i.e., conditioned on network geometry) signal-to-interference-plus-noise-ratio (SINR) performance metrics, which enables averaging via the moment generating function (MGF) of the interference. Although \cite[Sec.III]{survey_h}, \cite{Gil_marco, Rate_marco} show alternative techniques to circumvent the necessity of such representation, there are yet no SG models including state-of-the-art fading models that may include individually resolvable multipath components, which are denoted as multiple {\em dominant specular components} (DSCs)~\cite{Norman_MSC}. As argued in \cite{Norman_MSC, Yacoub,  Rapaport, GTR_just1, Slim_GTW}, both the Rayleigh and Nakagami-m models may fail to capture realistic fading environments. The Rayleigh fading ignores the line-of-sight (LOS) component in the received signal, which is prominent in outdoor cellular communication. It is also a single parameter fading model that is not flexible enough to model complex indoor fading environment. The Nakagami-m fading can neither aptly capture fading with LOS component nor accurately model indoor fading environment as argued in \cite{Norman_MSC, Yacoub, GTR_just1, Rapaport}. Furthermore, the tractability of SG analysis with Nakagami-m fading necessities an integer value for $m$, which decreases the model's flexibility to capture complex fading environments.

In order to develop SG models in terms of multipath fading, it is required to incorporate a flexible fading model that can span different fading channels including those with multiple DSCs. The state-of-the-art studies in multi-path fading (see  \cite{Norman_MSC, Yacoub,  Rapaport, GTR_just1, Slim_GTW} and the references therein)  show that fading channels with a finite number of DSCs with dominant powers in addition to a diffuse component appear in several practical indoor/outdoor scenarios. A simple example with single DSC appears cellular networks where one LOS path exists between the BS and a mobile user. In this case, the resulting fading environment follows the well-known Rician fading model. Furthermore, multiple DSCs may be created by reflections from metal objects (e.g., light-posts, cars) in close proximity to the BSs and/or user, in which the reflections would have comparable amplitude to the dominant LOS path due to the negligible power absorption factor of metals. A fading channel with a small number of specular components without a diffuse component may appear in millimeter wave (mmW) communication where high directional antennas are used for short-range communication~\cite{Rappaport_mmW}. Also, severe NLOS fading environments can be modeled via a small number of specular components without diffuse components~\cite{Slim_GTW, Rapaport, GTR_just1}. Last but not least, field measurements for indoor multipath fading channels confirm the existence of multiple DSCs in addition to the diffuse component~\cite{Rapaport2}. Therefore, there is a need to incorporate new flexible fading models that capture realistic multipath fading environments with multiple DSCs.

As a step forward to account for fading channels with multiple DSCs, this paper incorporates the generalized two-ray (GTR) fading model into tractable stochastic geometry analysis. The GTR model is adopted due to its mathematical elegance and practical significance\footnote{Models with more than two dominant specular components are not likely to appear in practice~\cite{Slim_GTW, Rapaport}.}. The GTR fading channel is characterized by four tunable parameters that explicitly account for two DSCs plus the diffuse component, which gives high flexibility to capture a diverse spectrum of fading channels. With the proper adjustment of the GTR parameters, it captures several fading channels as special cases such as {\em deterministic}, {\em Rician}, {\em Rayleigh}, {\em hyper-two ray}, and {\em hyper-Rayleigh}~\cite{Slim_GTW, GTR_just1}. Hence, accounting for GTR fading in a SG model leads to unified analysis for a diverse spectrum of fading channels. 
It is worth noting that the two DSCs, which are explicitly accounted for by the GTR model, are not necessarily LOS components. However, they may also be considered as two NLOS paths that experienced favorable propagation when compared to the others in the diffuse part. Using this interpretation, the severe multi-path fading channels captured by the GTR model can be justified \cite{Slim_GTW, GTR_just1}.

To the best of the authors' knowledge, this work is the first to incorporate such a flexible fading model into a tractable SG analysis. It is worth mentioning that we incorporate the GTR fading model into a simplistic network setup to observe its explicit effects. Analyzing more advanced network setups with GTR fading is postponed to future work.

\section{System Model}\label{sec:Sys_Model}

We consider a downlink single-tier cellular network with single-antenna BSs that are deployed according to the PPP $\Psi$ with intensity $\lambda$. Each element $x_i  \in \Psi$ belongs to $\mathbb{R}^2$ and denotes the location of the i$^{th}$ BS. Single-antenna users' equipment (UEs) are spatially distributed according to a stationary point process $\Phi$ with intensity $\mathcal{U}$ such that $\mathcal{U} \gg \lambda$. All BSs transmit with a constant transmit power of P. Users associate to the BSs according to the average radio signal strength (RSS) rule.  Universal frequency reuse is adopted with no intra-cell interference.

It is assumed that the signal power decays according to the power-law $r^{-\eta}$ with the distance $r$, where $\eta$ is the path-loss exponent. We focus on the performance of a test UE located at the origin. According to Slivnyak's theorem, there is no loss of generality with  this assumption. For the sake of simple exposition, we define the set $\tilde{\Psi} \in \mathbb{R}$ that contains the ordered distances from the test user to the BSs in $\Psi$. Following the RSS rule, $r_0$ and $r_i$ are the distance between the test UE and, respectively, his serving BS and the i$^{th}$ nearest interfering BS.

\subsection{The GTR Fading Model}

We consider a GTR model with two specular components plus a diffuse component. The received baseband signal can be represented as:

\small
\begin{align} \label{signal}
& \!\!\!\!\!\!\!\!\!\!\!  y(r)=\sqrt{\frac{P}{r_0^\eta}} \underset{g_0}{\underbrace{\left(V_{\{0,1\}} e^{j \phi_{\{0,1\}}} + V_{\{0,2\}} e^{j \phi_{\{0,2\}}} + X_0+jY_0\right)}} s_0 \notag \\
& \!\!\!\!\!\!\!\!\!\!\! \!\!\!\! +\!\!\!\!   \sum_{r_i \in \tilde{\Psi}\setminus r_0}\sqrt{\frac{P}{r_i^\eta}} \underset{g_i}{\underbrace{\left(V_{\{i,1\}} e^{j \phi_{\{i,1\}}} + V_{\{i,2\}} e^{j \phi_{\{i,2\}}} + X_i+jY_i\right)}} s_i+ n,
\end{align}\normalsize
where $s_0$ and $s_i$ are respectively the intended and i$^{th}$ BS interfering symbols which are assumed to be drawn from Gaussian codebooks, $(V_{0,1}, \phi_{0,1})$ and $(V_{0,2}, \phi_{0,2})$ are the ({\em amplitude, phase}) of the two DSCs of the intended signal, $X_0\sim\mathcal{N}(0,\sigma_0^2)$ and $Y_0\sim\mathcal{N}(0,\sigma_0^2)$ are the zero mean Gaussian distributed intended signal diffuse components with total power of $2 \sigma_0^2$, $(V_{i,1}, \phi_{i,1})$ and $(V_{i,2}, \phi_{i,2})$ are the ({\em amplitude, phase}) of the two DSCs of the interfering signal from the i$^{th}$ BS, $X_i\sim\mathcal{N}(0,\sigma_i^2)$ and $Y_i\sim\mathcal{N}(0,\sigma_i^2)$ are the zero mean Gaussian distributed i$^{th}$ BS interfering signal diffuse components with total power of $2 \sigma_i^2$, and $n\sim\mathcal{CN}(0,\frac{N_0}{2})$ is the complex Gaussian noise. Let $\alpha_i = \phi_{i,1} - \phi_{i,2}$ be the phase difference between the two DSCs and  $f_{\alpha_i}(.)$ denotes the probability density function (PDF) of $\alpha_i$, then according to \cite{Slim_GTW}, the GTR model is defined by four parameters, namely the diffuse power (DP) $2 \sigma_i^2$, the specular-component-to-diffuse-power-ratio (SDPR)  $K_i = \frac{V_{i,1}^2+V_{i,2}^2}{2 \sigma_i^2}$, peak-to-average-specular-component-power-ratio (PASPR) $\Delta_i =  \frac{2 V_{i,1} V_{i,2}}{V_{i,1}^2 +V_{i,2}^2}$, and $f_{\alpha_i}(.)$.

Observing $K_i$ and $\Delta_i$, we notice that the GTR model reduces to Rayleigh fading for $K=0$ and reduces to Rician fading for $\Delta = 0$.  It also captures the two-ray, hyper-two ray, and hyper-Rayleigh by letting $K \rightarrow \infty$, $\sigma^2 \rightarrow 0$, and properly choosing $f_{\alpha}(.)$.  The distribution function $f_{\alpha}(.)$ discriminates between different types of GTR fading models.  Particularly, $\alpha$ is uniformly distributed in the range of  $[0,2 \pi]$ in the GTR-U model, whereas $\alpha$ is uniformly distributed in the truncated range of  $[\pi(1-p),\pi(1+p)]$ in the GTR-T model. In the GTR-V model, $\alpha$ follows the Von Mises distribution. Note that the GTR-T and GTR-V capture the cases when the angles of the two specular components are correlated, which appears in some practical scenarios~\cite{Slim_GTW}.

\subsection{Methodology of Analysis} \label{method}\label{sec:Meth}

For a given realization of network geometry and channels' gains, the interference term in \eqref{signal} has a Gaussian distribution. Hence, treating interference as noise, the instantaneous SINR for the test user can be expressed as:

\begin{align}\label{eq:SINR}
{\rm SINR}(r_0)&=\frac{P |g_o|^2 r_o^{-\eta}}{{{\sum\limits_{i \in \tilde{\Psi}\setminus r_0}P |g_i|^2 r_i^{-\eta}}}+N_o}
&=\frac{\mathcal{S}(r_0)}{\mathcal{I}(r_0)+N_o},
\end{align}
where $r_0$ is excluded from $\tilde{\Psi}$ in \eqref{eq:SINR} because the serving BS does not contribute to the interference. In SG analysis, we are interested in spatially averaged performance metrics, which requires the conditional (i.e., conditioning on $r_0$) PDFs of $\mathcal{S}(r_0)$ and $\mathcal{I}(r_0)$, in addition to the PDF of $r_0$.  In a PPP network with RSS association, the PDF of $r_0$ is known to be $f_R(r_o)=2 \pi \lambda r_o \exp(- \pi \lambda r_o^2)$, $0\leq r_o<\infty$ \cite{tractable_app}. Also, the conditional PDF of $\mathcal{S}(r_0)$ is straightforward to obtain from the PDF of $|g_0|^2$. However, the aggregate interference $\mathcal{I}(r_0)$ is usually characterized via its Laplace transform (LT)\footnote{With slight abuse of notation,  LT is used to denote the LT of a probability density function (PDF) of a random variable, which is equivalent to the moment generating function with negative argument. }.  Hence, only performance metrics that are expressed in terms of the LT of $\mathcal{I}(r_0)$ can be evaluated.

\begin{figure*}
\begin{center}
\small
\begin{align}\label{eq:GTR_Exact}
\!\!\!\!\!\!\!\!\!\!\!\! \mathcal{L}_{\mathcal{I}(r_o)}(s)   &{=} \exp \Bigg(\pi \lambda   \left(\frac{2 \sigma^2 s P (1-y_o)}{y_o }\right)^\frac{2}{\eta} \left(1-  (1-y_o) \zeta_1 \left(K y_o \right) \right) \notag \\
&\quad \quad \quad \quad \quad \quad\quad\quad\quad\quad\quad\quad\quad\quad\quad\quad - \pi \lambda \left(2 \sigma^2 s P\right)^\frac{2}{\eta}  \left[\int_0^{y_0}  \frac{(1-y)^{\frac{2}{\eta}+1} }{y^\frac{2}{\eta} }  \zeta_1^{'} \left(K y \right)  dy  + \int_0^{y_0}   \frac{(1-y)^{\frac{2}{\eta}}\zeta_1 \left(K y \right) }{y^\frac{2}{\eta} }  dy \right] \Bigg).
\end{align}
\hrulefill
\small
\setcounter{equation}{5}
\begin{align}\label{eq:GTR_LB}
\!\!\!\!\!\!\!\!\!\! \underline{\mathcal{L}_{\mathcal{I}(r_o)}}(s)   &{=} \exp \Bigg(\pi \lambda   \left(\frac{2 \sigma^2 s P (1-y_o)}{y_o }\right)^\frac{2}{\eta} \left(1-  (1-y_o) \zeta_1 \left(K y_o \right) \right) - \frac{2 \pi^2}{\eta} \zeta_2 \lambda \csc \left(\frac{2 \pi }{\eta} \right) \left(2 \sigma^2 s P\right)^\frac{2}{\eta} \Bigg).
\end{align}
\hrulefill
\end{center}
\end{figure*}

For the sake of organized presentation, we devote Section~\ref{sec:interference} to derive the conditional LT of the aggregate interference with different GTR fading environments. In Section~\ref{sec:analysis}, we first  express  the averaged spectral efficiency (i.e., $\mathbb{E}\left[\ln\left(1+{\rm SINR}(r_0) \right)\right]$) in terms of the conditional LTs derived in Section~\ref{sec:interference}. Then, the averaging step over $r_0$ is done to obtain the spatially average spectral efficiency.

\section{Interference Characterization in GTR Fading Environment}\label{sec:interference}

We assume that all fading channels are independent from the BSs' locations, independent from each other, and are identically distributed (i.i.d.) according to the GTR model with parameters: $K$, $\Delta$, $\sigma^2$, and $f_{\alpha} (.)$. Let $\Omega=V_1^2 + V_2^2+ 2 \sigma^2$ represents the expected power gain of the fading channel, then from the definition of $K$,  we have $\Omega=(K+1)2 \sigma^2$. For general $f_{\alpha}(.)$, the LT of the aggregate interference in GTR fading environment is  given by the following lemma.

\normalsize
\begin{lemma}

In a GTR fading environment with parameters $K$, $\Delta$, $\sigma^2$, and $f_{\alpha}(.)$, the LT of the conditional aggregate interference (conditional on $r_o$) in a PPP cellular network with intensity $\lambda$ is given by equation \eqref{eq:GTR_Exact}, where $y_o(s)= \frac{  2 \sigma^2s  P}{r_o^{\eta} +2 \sigma^2 s P}$, and

\setcounter{equation}{3}
\small
\begin{align}\label{eq::GeneralZ1}
&\zeta_1 \left(y \right) = \int\limits_{0}^{2 \pi} e^{-y (1+\Delta \cos(\alpha))} f_{\alpha} \left({\alpha}\right) d {\alpha},  \\
&\zeta_1' \left(Ky \right) = \frac{\zeta_1 \left(Ky \right)}{dy}.
\end{align}
\normalsize

\begin{proof}
Refer to Appendix \ref{sec:AppA}.
\end{proof}
\end{lemma}

The expression in \eqref{eq:GTR_Exact} contains two integrals, which may increase the computational complexity. Therefore, we obtain a simpler lower-bound on \eqref{eq:GTR_Exact} which is given by the following lemma.

\setcounter{equation}{6}
\normalsize
\begin{lemma} \label{LB_Ricain}
The LT given in equation \eqref{eq:GTR_Exact} is lower bounded by \eqref{eq:GTR_LB}, where ${}_1 F_1(.)$ is the confluent hypergeometric function, $\zeta_1(.)$ is given by equation \eqref{eq::GeneralZ1}, and

\small
\begin{align}
&\!\!\!\!\!\!\!\!\!\!\!\! \zeta_2  = \int\limits_{0}^{2 \pi}f_{\alpha} \left({\alpha}\right) \Bigg[ {}_1 F_1 \left(1- \frac{2}{\eta};2;-K(1+\Delta \cos (\alpha)) \right)  + \notag \\
&\!\!\!\!\!\!\!\!\!\!\!\! K(1+\Delta \cos (\alpha)) \frac{\eta+2}{2 \eta}  {}_1 F_1 \left(1- \frac{2}{\eta};3;-K(1+\Delta \cos (\alpha)) \right) \Bigg]  d {\alpha}.
\end{align}
\normalsize
\begin{proof}
Refer to Appendix \ref{sec:AppB}.
\end{proof}
\end{lemma}

\normalsize
It is worth emphasizing that $\zeta_2$ is independent of the LT variable and is a function of the fading parameters only. Hence, for a known fading parameters, $\zeta_2$ is a constant w.r.t. the interference LT. The accuracy of the lower-bound in \eqref{eq:GTR_LB} is validated in Section~\ref{sec:Results}. Based on Lemma 1 and Lemma 2, the interference for different GTR fading models can be characterized. For instance, the interference with Rician  faded channels is characterized by the following corollary,

\begin{corollary}

In a Rician fading environment with parameters $K$ and $\sigma^2$, the LT of the conditional aggregate interference in a PPP cellular network with intensity $\lambda$ is given by equation \eqref{eq:GTR_Exact}, and lower-bounded by \eqref{eq:GTR_LB}, where $y_o(s)= \frac{  2 \sigma^2s P }{r_o^{\eta} +2 \sigma^2 s P  }$ and,

\small
\begin{align}
&\!\!\!\!\!\!\!\!\!\!\!\! \zeta_1(x)=e^{-x},
\end{align}
\begin{align}
&\!\!\!\!\!\!\!\!\!\!\!\! \zeta_2  =  K \frac{\eta+2}{2 \eta}  {}_1 F_1 \left(1- \frac{2}{\eta};3;-K\right)  +  {}_1 F_1 \left(1- \frac{2}{\eta};2;-K \right) .
\end{align}
\normalsize
\begin{proof}
The corollary is obtained from  Lemma 1 and Lemma 2 by setting $\Delta=0$.
\end{proof}
\end{corollary}

Interestingly, $\zeta_1(.)$ and $\zeta_2$ for Rician  fading is obtained in closed forms. The interference in GTR-U fading case is characterized via the following corollary,
\begin{corollary}

In a GTR-U fading environment with parameters $K$, $\Delta$, and $\sigma^2$, the LT of the conditional aggregate interference in a PPP cellular network with intensity $\lambda$ is given by equation \eqref{eq:GTR_Exact}, and lower-bounded by \eqref{eq:GTR_LB}, where $y_o(s)= \frac{  2 \sigma^2s  P}{r_o^{\eta} +2 \sigma^2 s   P}$ and,

\small
\begin{align}
&\!\!\!\!\!\!\!\!\!\!\!\! \zeta_1(x)=e^{-x} I_o \left(  x \Delta \right),
\end{align}
\begin{align}
&\!\!\!\!\!\!\!\!\!\!\!\! \zeta_2  = \int\limits_{0}^{2 \pi} \Bigg[{}_1 F_1 \left(1- \frac{2}{\eta};2;-K(1+\Delta \cos (\alpha)) \right)   + \notag \\
&\!\!\!\!\!\!\!\!\!\!\!\! K(1+\Delta \cos (\alpha)) \frac{\eta+2}{2 \eta}  {}_1 F_1 \left(1- \frac{2}{\eta};3;-K(1+\Delta \cos (\alpha)) \right) \Bigg] \frac{d {\alpha}}{2 \pi}.
\end{align}
\normalsize
\begin{proof}
The corollary is obtained from  Lemma 1 and Lemma 2 by setting $f_{\alpha}(\alpha)= \frac{1}{2 \pi}$.
\end{proof}
\end{corollary}

The GTR-U assumes that the phases of the two DSCs (i.e., $\phi_1$ and $\phi_2$) are independent, and hence, the value of the phase difference $\alpha$ is uniformly distributed from $[0,2 \pi]$. In some cases, correlation between $\phi_1$ and $\phi_2$  may exist, which limits the range that $\alpha$ spans. The scenarios where $\phi_1$ and $\phi_2$ are correlated are captured via the GTR-T and GTR-V cases. In the GTR-T, the distribution of $\alpha$ is assumed to be uniformly distributed in the range of $[\pi(1-p),\pi(1+p)]$, where $p \in [0,1]$ is the truncation parameter that can be manipulated to capture different correlation scenarios. In the GTR-V case, $\alpha$ is assumed to follow the Von Mises, given by
\begin{align} \label{VonM}
f_{\alpha} \left( \alpha \right)= \frac{\exp \left(-\delta \cos \left(\alpha \right) \right)}{2 \pi I_o \left(\delta \right)},\ \ \ \ \alpha \in [0,2 \pi]
\end{align}
where $\delta$ is the phase difference distribution parameter that can be manipulated to concentrate the PDF of $\alpha$ around a certain value in the $[0,2 \pi]$ range. It is worth mentioning that both the GTR-T and GTR-V reduce to the GTR-U for $p=1$ and $\delta=0$, respectively. The LT of the conditional aggregate interference in the cases of GTR-T and GTR-V is given by the following corollaries

\begin{corollary}

In a GTR-T fading environment with parameters $K$, $\Delta$, $\sigma^2$, and $p$, the LT of the conditional aggregate interference in a PPP cellular network with intensity $\lambda$ is given by equation \eqref{eq:GTR_Exact}, and lower-bounded by \eqref{eq:GTR_LB}, where $y_o(s)= \frac{  2 \sigma^2s P }{r_o^{\eta} +2 \sigma^2 s P  }$ and,

\small
\begin{align}
\!\!\!\!\!\!\!\!\!\!\!\! \zeta_1 \left(x \right) =\int\limits_{\pi (1-p)}^{\pi (1+p)} \frac{e^{-x (1+\Delta \cos(\alpha))}}{{2 \pi  p }}  d {\alpha},
\end{align}
\begin{align}
&\!\!\!\!\!\!\!\!\!\!\!\! \zeta_2  =\int\limits_{\pi (1-p)}^{\pi (1+p)}\frac{d {\alpha}}{2 p \pi} \Bigg[ {}_1 F_1 \left(1- \frac{2}{\eta};2;-K(1+\Delta \cos (\alpha)) \right)   + \notag \\
&\!\!\!\!\!\!\!\!\!\!\!\! \frac{ K(1+\Delta \cos (\alpha))(\eta+2)}{2 \eta}  {}_1 F_1 \left(1- \frac{2}{\eta};3;-K(1+\Delta \cos (\alpha)) \right) \Bigg] .
\end{align}
\normalsize
\begin{proof}
The corollary is obtained from  Lemma 1 and Lemma 2 by setting $f_{\alpha}(\alpha)= \frac{1}{2 \pi p}$ for $\pi(1-p) < \alpha < \pi(1+p)$, and $f_{\alpha}(\alpha)=0$ otherwise.
\end{proof}
\end{corollary}

\begin{corollary}

In a GTR-V fading environment with parameters $K$, $\Delta$, $\sigma^2$, and $\delta$, the LT of the conditional aggregate interference in a PPP cellular network with intensity $\lambda$ is given by equation \eqref{eq:GTR_Exact}, and lower-bounded by \eqref{eq:GTR_LB}, where $y_o(s)= \frac{  2 \sigma^2s P }{r_o^{\eta} +2 \sigma^2 s P  }$ and,

\small
\begin{align}
&\!\!\!\!\!\!\!\!\!\!\!\! \zeta_1(x)= \frac{I_o \left(  x \Delta+ \delta \right) e^{- x}}{I_o \left(\delta \right)}.
\end{align}
\normalsize
\small
\begin{align}
&\!\!\!\!\!\!\!\!\!\!\!\! \zeta_2  = \int\limits_{0}^{2 \pi} \frac{\exp \left(-\delta \cos \left(\alpha \right) \right)}{2 \pi I_o \left(\delta \right)} \Bigg[{}_1 F_1 \left(1- \frac{2}{\eta};2;-K(1+\Delta \cos (\alpha)) \right)   + \notag \\
&\!\!\!\!\!\!\!\!\!\!\!\! K(1+\Delta \cos (\alpha)) \frac{\eta+2}{2 \eta}  {}_1 F_1 \left(1- \frac{2}{\eta};3;-K(1+\Delta \cos (\alpha)) \right) \Bigg]   d {\alpha}.
\end{align}
\normalsize

\begin{proof}
The corollary is obtained from  Lemma 1 and Lemma 2 by using \eqref{VonM} for the distribution of $\alpha$.
\end{proof}
\end{corollary}

The cases provided in Lemma 1, Lemma 2, and Corollaries 1 to 4 represent the case in which the DSCs arrive in addition to a diffuse component. This implicitly implies that interference signals have experienced favorable fading channels. Another case of special interest is when interference links experience severe fading, which can be captured via a two ray model without a diffuse component~\cite{Slim_GTW}. In this case, we set $K \rightarrow \infty$ and $2 \sigma^2 \rightarrow 0$ such that $\Omega = (K+1) 2 \sigma^2$ is kept constant. For a general $f_{\alpha}(\alpha)$ the LT of the aggregated interference in severe fading environment is given in the following lemma,

\begin{lemma}
The LT of the aggregate interference in a GTR fading environment with $K \rightarrow \infty$ and $\sigma^2 \rightarrow 0$ such that $(K+1) 2 \sigma^2$ is constant, is given by

\small
\begin{align}\label{eq:GTR_approx}
\!\!\!\!\!\!\!\!\!\! \mathcal{L}_{{\mathcal{I}(r_o)}}(s) {=}  \exp \Bigg(\pi \lambda \left(r_o^2 \zeta_1(s P \Omega r_o^{-\eta}) -(s P)^{\frac{2}{\eta}} {\zeta_3}(s P \Omega r_o^{-\eta})\right) \Bigg),
\end{align}\normalsize
such that,
\small
\begin{align}\label{eq:ZetaBar_Gen}
&\!\!\!\!\!\!\!\!\!\!\!\! {\zeta_3}(x)  = \int\limits_{0}^{2 \pi} \gamma \left( 1-\frac{2}{\eta},x (1+\Delta \cos (\alpha))  \right)  f_{\alpha} \left({\alpha}\right) d {\alpha}.
\end{align}\normalsize
where $\gamma(.,.)$ denotes the lower incomplete gamma function and $\zeta_1(.)$ is given by equation \eqref{eq::GeneralZ1}.
\begin{proof}
Refer to Appendix C.
\end{proof}
\end{lemma}

For each of the special cases of GTR fading, ${\zeta_3}(.)$ in \eqref{eq:ZetaBar_Gen} can be evaluated via the appropriate $f_{\alpha}(.)$. For GTR-U, GTR-T, and GTR-V, expressions for ${\zeta_3}(.)$ are given, respectively, by

\small
\begin{align}
&\!\!\!\!\!\!\!\!\!\!\!\! {\zeta_3}(x)  = \int\limits_{0}^{2 \pi} \gamma \left( 1-\frac{2}{\eta},x(1+\Delta \cos (\alpha)) \right)  \frac{d {\alpha}}{2 \pi}.\\
&\!\!\!\!\!\!\!\!\!\!\!\! {\zeta_3}(x)  = \int\limits_{\pi (1-p)}^{\pi (1+p)} \gamma \left( 1-\frac{2}{\eta},x(1+\Delta \cos (\alpha)) \right)  \frac{d {\alpha}}{2 p \pi}. \\
&\!\!\!\!\!\!\!\!\!\!\!\! {\zeta_3}(x)  = \int\limits_{0}^{2 \pi} \gamma \left( 1-\frac{2}{\eta},x(1+\Delta \cos (\alpha)) \right)   \frac{\exp \left(-\delta \cos \left(\alpha \right) \right)}{2 \pi I_o \left(\delta \right)} d {\alpha}.
\end{align}\normalsize

\setcounter{equation}{21}
\begin{figure*}
\small
\begin{equation} \label{eq:useful1}
\mathcal{L}_{\mathcal{S}(r_o)}(s) = \frac{1}{1+2 \sigma^2 s \bar{P}(r_0)}\exp\left(\frac{-K 2 \sigma^2 s\bar{P}(r_0)}{1+2 \sigma^2 s \bar{P}(r_0)}\right) I_o\left(\frac{-\Delta K 2 \sigma^2s\bar{P}(r_0)}{1+s 2 \sigma^2\bar{P}(r_0)} \right).
\end{equation}
\hrulefill
\begin{equation}\label{eq:useful2}
\mathcal{L}_{\mathcal{S}(r_o)}(s) = \frac{1}{2 \pi p} \int_{\pi (1-p)}^{\pi (1+p)}\frac{1}{1+s 2 \sigma^2  \bar{P}(r_0)}\exp\left(\frac{-K (1+\Delta \cos(\alpha)) 2 \sigma^2 s \bar{P}(r_0)}{1+s 2 \sigma^2  \bar{P}(r_0)}\right)  d\alpha.
\end{equation}
\hrulefill
\begin{equation}\label{eq:useful3}
\mathcal{L}_{\mathcal{S}(r_o)}(s) = \frac{1}{I_o(\delta)(1+s 2 \sigma^2  \bar{P}(r_0))}\exp\left(\frac{-K 2 \sigma^2 s \bar{P}(r_0)}{1+s 2 \sigma^2  \bar{P}(r_0)}\right) I_o\left(\delta-\frac{\Delta K 2 \sigma^2 s \bar{P}(r_0)}{1+2 \sigma^2 s  \bar{P}(r_0)} \right).
\end{equation}
\hrulefill
\normalsize
\end{figure*}

It is worth highlighting that the severe fading channels are captured by the constructive and destructive interference of the two rays. Consequently, the two DSCs should be explicitly consider to account for such severe fading cases. In contrast, the Rician  fading accounts only for one specular component plus the diffuse component, and hence, setting $K \rightarrow \infty$ and $\sigma^2 \rightarrow 0$ leads to a deterministic channel gain. 

\section{Performance Analysis}\label{sec:analysis}
In this section, the conditional LTs of the aggregate interference obtained in Section~\ref{sec:interference} are utilized to evaluate the spatially averaged  spectral efficiency.

\subsection{Average Spectral Efficiency}


Starting form the Lemma introduced in \cite{hamdi2010useful}, which states that

\setcounter{equation}{20}
\small
\begin{align} \label{hamdis}
 & \!\!\!\!\!\!\!\!\!\!\!\!\! \mathbb{E}\left[\ln\left(1+\frac{X}{Y+1}\right) \right] = \int_0^\infty \frac{\mathcal{L}_{Y}(z)-\mathcal{L}_{X,Y}(z)}{z} \exp\{-z\} dz.
 \end{align}
\normalsize
where $\mathcal{L}_{X,Y}(z) = \mathcal{L}_{X}(z) \mathcal{L}_{Y}(z)$.

Exploiting \eqref{hamdis} the spectral efficiency can be directly expressed in terms of the LT of interfering and useful links. Following \cite{Slim_GTW}, the LT of the desired signal power while conditioning on $r_o$ is given by equations \eqref{eq:useful1}, \eqref{eq:useful2}, and \eqref{eq:useful3} for GTR-U, GTR-T, and GTR-V fading channel, respectively, where $\bar{P}(r_0)=P r_o^{-\eta}$.

\begin{theorem}
The average spectral efficiency for a PPP cellular network with intensity $\lambda$ in a GTR fading environment with parameters $\left\{K_o,\Delta_o,\sigma_o^2,f_{\alpha_o}(.)\right\}$ and $\left\{K,\Delta,\sigma^2,f_{\alpha}(.)\right\}$ for the desired and interfering fading channels is given by the following equation

\setcounter{equation}{24}
\small
\begin{align}
& \!\!\!\!\!\!\!\!\!\!\!\! \mathcal{\bar{A}} = \notag \\
 & \!\!\!\!\!\!\!\!\!\!\!\! \int_0^\infty \int_0^\infty \mathcal{L}_{\mathcal{I}(r_o)}\left(\frac{z}{N_o}\right)\left(1-\mathcal{L}_{\mathcal{S}(r_o)}\left(\frac{z}{N_o}\right)\right) \frac{e^{-z}}{z} f_{r_0}(r) dr dz.
\end{align}\normalsize
where $ \mathcal{L}_{\mathcal{I}}(.)$ is given by equation \eqref{eq:GTR_Exact} for a general $K$, lower bounded by equation \eqref{eq:GTR_LB} and when $K \rightarrow \infty$ it reduces to equation \eqref{eq:GTR_approx}.
\begin{proof}
Follows from substituting \eqref{eq:SINR} in \eqref{hamdis} and averaging over $r_o$.
\end{proof}
\end{theorem}

\section{Results}\label{sec:Results}

\begin{figure}[t]
\centerline{\includegraphics[width=  3in]{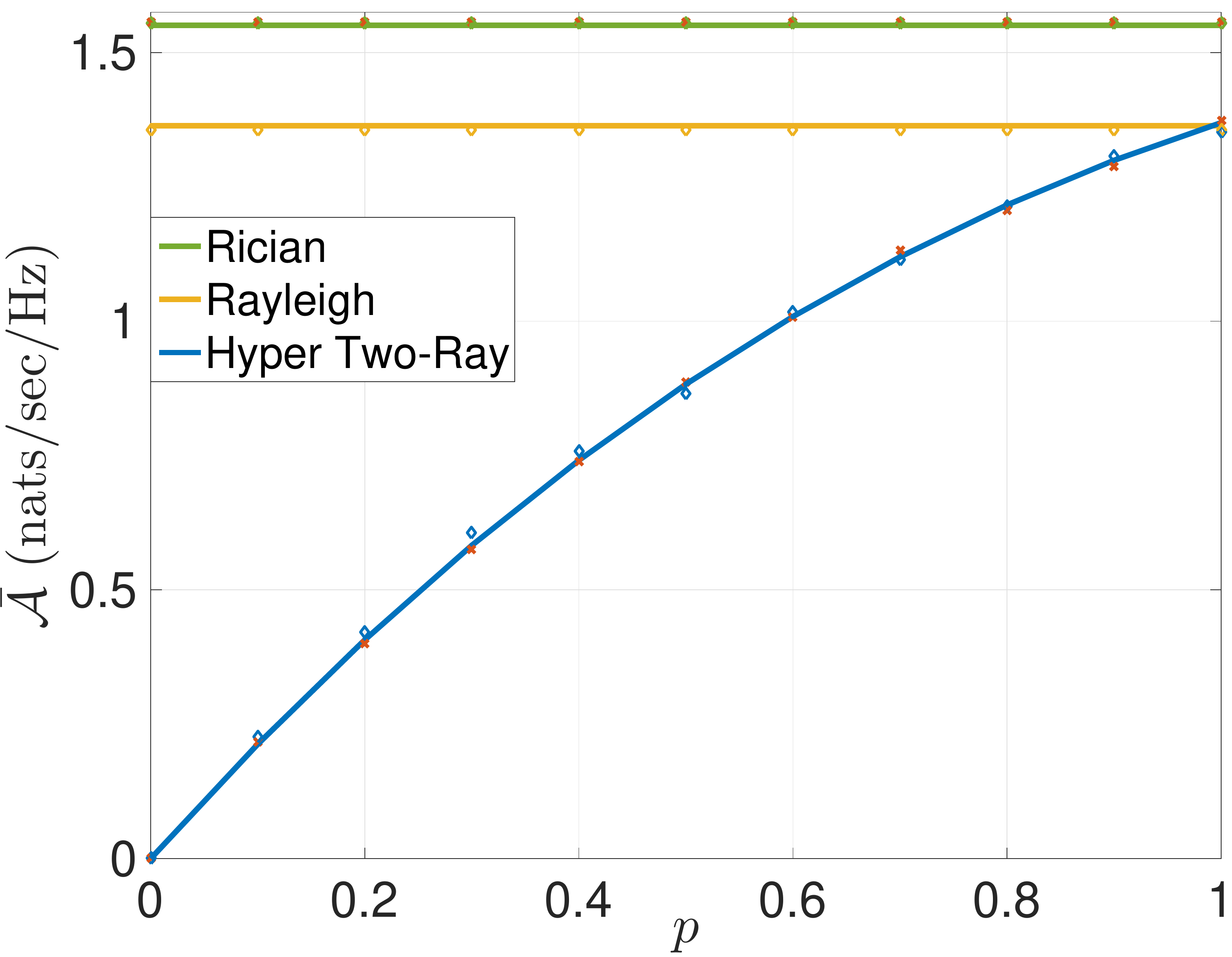}}
\caption{\,\small Average spectral efficiency vs. truncation parameter (p) assuming GTR-T, Rician and Rayleigh for the desired channel and GTR-U for interfering channels. Solid lines, the diamonds and the X's represent the results obtained analytically, by simulations and by simulations assuming Rayleigh fading for all the fading channels except the nearest 5 BSs. The chosen values are the same for the two ray model where $K \rightarrow \infty$, $\Delta=1$, and $\sigma^2 \rightarrow 0$ such that the average received power is kept constant for all cases.}
\label{fig:Fig3}
\end{figure}

This section presents numerical results for the spectral efficiency for GTR fading environments. All results are validated by independent system level simulations. The match between the analysis and simulation of all cases confirms the validity of our expressions. Unless otherwise stated, we used the following parameters' values: $N_o=-80 \  {\rm dBm}$, $\lambda=3 \ {\rm BSs/Km}^2$, $P=3 \ {\rm W}$, $\Delta=1$, $\sigma^2=1$, and $\Omega_o=2$.

Fig.~\ref{fig:Fig3} shows the spectral efficiency obtained via Theorem 2 for GTR-U fading on the interfering links and different fading models on the useful link. For the sake of fair comparison, we keep a constant $\Omega$ for both links. For GTR-T on the useful link, the performance is highly affected by the value of $p$. Lower $p$ implies higher correlations between the DSCs phases which lead to a high probability of destructive interference, and vice versa. The figure also shows that Rician fading on the useful link has higher performance than the Rayleigh fading due to the LOS path, and higher performance than the GTR-T fading due to the absence of destructive DSCs interference.

\begin{figure}[t]
\centerline{\includegraphics[width=  3in]{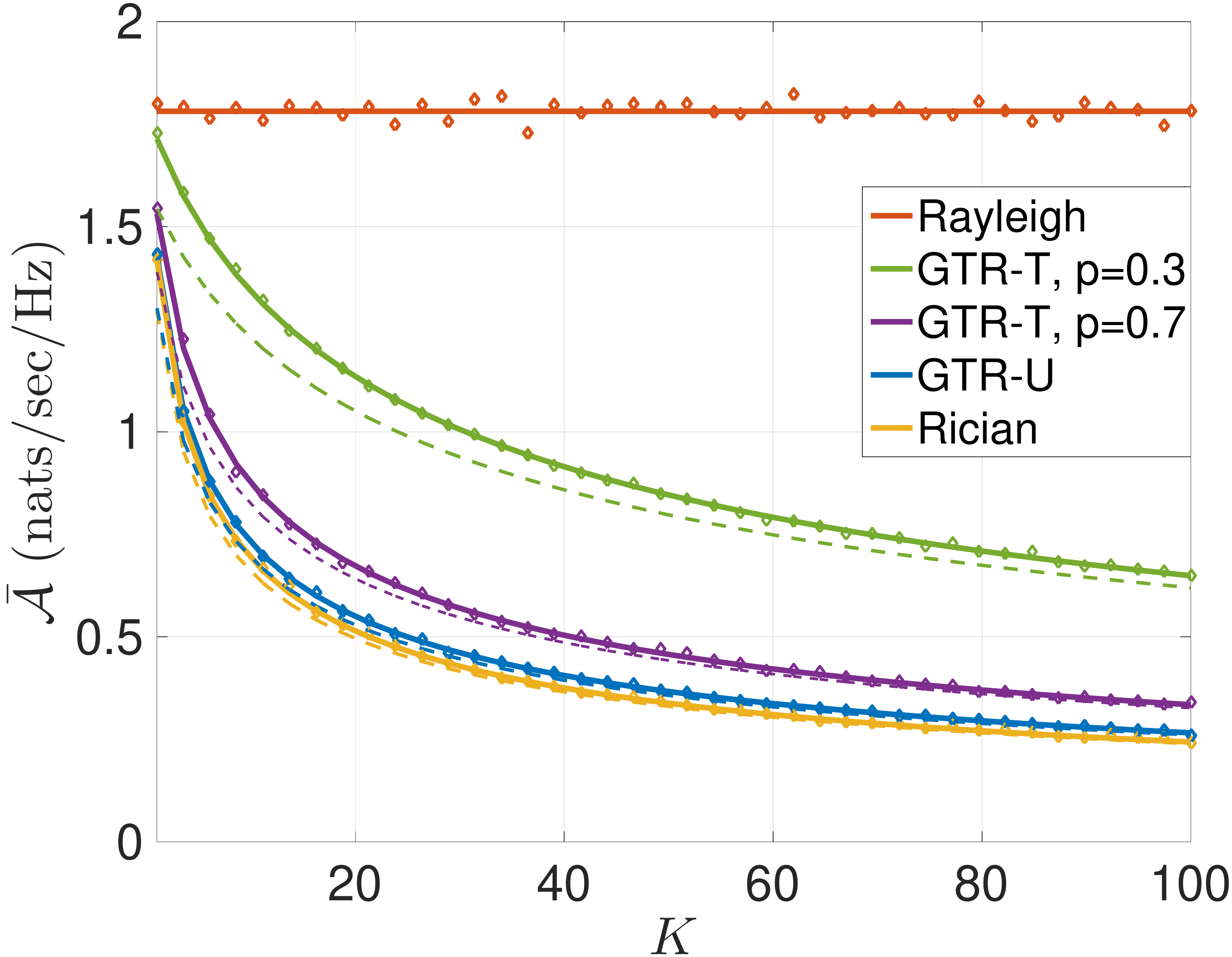}}
\caption{\,\small Ergodic rate assuming Rayleigh fading for the desired channel and GTR-T for the interferes channel. Solid lines, dashed lines, and the diamonds represent the rate obtained by the exact expression, the lower bound, and simulations.}
\label{fig:Fig1}
\end{figure}

In Fig.~\ref{fig:Fig1}, we show the explicit effect of fading on the interfering links by fixing GTR-U fading on the useful link. Note that we fix the diffuse power $2 \sigma^2$ for both the useful and interfering channels. In consistence with the previous results, Fig.\ref{fig:Fig1} manifests the effect of the interference fading model on the spectral efficiency. The Rayleigh fading contains the diffuse power only, and hence, has the highest spectral efficiency. In contrast, the Rician  fading has the strongest channel gain due to the LOS connection, and hence, it represents the worst interference and lowest spectral efficiency. The GTR fading with multiple DSCs performance lies between the Rayleigh and Rician cases due to the probability of destructive DSCs interference. By decreasing (increasing) $p$, the GTR-T approaches the Rayleigh (Rician) fading performance due to the higher (lower) probability of destructive interference. Last but not least, the figure confirms the tightness of the lower bound obtained by Lemma 2.

\subsection{Discussion}

These results emphasize the prominent effect of multiple DSCs on the network performance in terms of spectral efficiency. The results also show the flexibility of GTR-T model to capture different DSCs destructive interference scenarios, in which the correlation between the phase differences is captured by manipulating $p$ to control the range that $\alpha$ spans.

While applying the GTR model to the useful link is quite intuitive, it might not be as much obvious why do we need to apply the GTR model to the interference links. To elaborate this point, we give the following justifications:

\begin{itemize}
\item The interference is dominated by nearby interferers that may have LOS channels with the receivers especially in outdoor communication with macro BSs. As shown in Fig. 1, assuming GTR-U fading for the nearest 5 BSs only and Rayleigh fading for the other interferes does not have a noticeable effect on the averaged spectral efficiency.
\item In suburban and rural areas with parks and open spaces, LOS interference is most likely to happen.
\item The developed model is not restricted to LOS communication and may also model NLOS communication in which one or two of the NLOS paths have dominant powers.
\item For indoor environments, the specular component that go through doors and windows may have dominant powers w.r.t. the components that penetrate through walls with high attenuation coefficients. Hence, interfering signal coming from adjacent rooms/corridors may have multiple DSCs.
\item Beside the solid physical foundation of the GTR model, it offers a mathematically elegant and flexible model that spans different LOS and NLOS fading environments. For instance, it can capture extreme fading environments such as  {\em Rayleigh},  the $\kappa-\mu$, {\em hyper-Rayleigh} and other channels~\cite{Slim_GTW, GTR_just1}.
\end{itemize}

It is worth highlighting that the developed model parameters can be tuned to be a function of the interfering BS distance. More particularly, the fading severity can be adjusted to increase with the propagation distance, which complies with practice. Such modification is postponed to future work.

\section{Conclusion}

This paper presents a tractable stochastic geometry (SG) model with GTR fading channels that explicitly accounts for two dominant specular components (DSCs). Exact and lower bounds for the spectral efficiency are obtained. Depending on the phase difference between the two DSCs at the receiver, constructive/destructive interference may occur leading to high variability in the fading channel gains. To this end, the explicit effect of GTR models on the interference and useful links are investigated.  The results confirm the prominent effect of DSCs fading models on the  network performance in terms of spectral efficiency. As special case of the GTR, we consider the GTR-T in which the phase correlation between the two DSCs is captured by controlling the range that the phase difference spans. Finally, the model flexibility to capture several fading conditions, ranging from deterministic and favorable Rician to severe hyper-Rayleigh and hyper-two ray, in large-scale cellular networks is also highlighted.

\appendix
Due to space constraints, only the outlines of the proofs are highlighted.
\subsection{Proof of Lemma 1}\label{sec:AppA}
Following the same methodology in \cite{tractable_app} and using the MGF of GTR fading in \cite{Slim_GTW}, the LT of $\mathcal{I}(r_0)$ is obtained as

\small
\begin{align}\label{eq:AppA_1}
 &\!\!\!\!\!\!\!\!\!\!\!\! \mathcal{L}_{\mathcal{I}}(s) {=}  \exp \left\{- 2 \pi \lambda \int_{r_0}^\infty \mathbb{E}_{\hat{K}}\left[1- \frac{r^{\eta}}{r^{\eta} + 2 \sigma^2 s P } e^{ \frac{ - \hat{K} 2 \sigma^2 P s  }{r^{\eta} +2 \sigma^2 s P }} \right] r dr \right\}.
\end{align}\normalsize
where $\hat{K}=K(1+\Delta \cos(\alpha))$. By change of variables, $y= \frac{  2 \sigma^2s  }{r^{\eta}  +2 \sigma^2 s P   }$, integration by parts, and some manipulations, \eqref{eq:GTR_Exact} in Lemma 1 is obtained.
\vspace{0.1cm}
\subsection{Proof of Lemma 2}\label{sec:AppB}
The integration in \eqref{eq:GTR_Exact} can be only evaluated when the upper limit is $1$. By definition, $y_o(s)$ is always less than one. Since the integrand in \eqref{eq:GTR_Exact}  is positive, setting the integral upper limit to $1$ and using  \cite[Eq.(3.383)]{integrals_book}, the lower bound in \eqref{eq:GTR_LB} is obtained.

\subsection{Proof of Lemma 3}\label{sec:AppC}
Setting $K \rightarrow \infty$ and $\sigma \rightarrow 0$ the average channel power is given by $\Omega= 2 \sigma^2 K$ Starting form \eqref{eq:AppA_1} and substituting $\hat{K}$ by $K (1+\Delta \cos (\alpha))$, $2 \sigma^2 K$ by $\Omega$, setting $\sigma^2 =0$. Then, by change of variables and integration by parts Lemma 3 is obtained.

\bibliographystyle{IEEEtran}
\bibliography{IEEEabrv,ref2}

\begin{thebibliography}{10}
\providecommand{\url}[1]{#1}
\csname url@samestyle\endcsname
\providecommand{\newblock}{\relax}
\providecommand{\bibinfo}[2]{#2}
\providecommand{\BIBentrySTDinterwordspacing}{\spaceskip=0pt\relax}
\providecommand{\BIBentryALTinterwordstretchfactor}{4}
\providecommand{\BIBentryALTinterwordspacing}{\spaceskip=\fontdimen2\font plus
\BIBentryALTinterwordstretchfactor\fontdimen3\font minus
  \fontdimen4\font\relax}
\providecommand{\BIBforeignlanguage}[2]{{%
\expandafter\ifx\csname l@#1\endcsname\relax
\typeout{** WARNING: IEEEtran.bst: No hyphenation pattern has been}%
\typeout{** loaded for the language `#1'. Using the pattern for}%
\typeout{** the default language instead.}%
\else
\language=\csname l@#1\endcsname
\fi
#2}}
\providecommand{\BIBdecl}{\relax}
\BIBdecl

\bibitem{tractable_app}
J.~G. Andrews, F.~Baccelli, and R.~K. Ganti, ``A tractable approach to coverage
  and rate in cellular networks,'' \emph{{IEEE} Trans. Commun.}, vol.~59,
  no.~11, pp. 3122--3134, Nov. 2011.

\bibitem{martin_ppp}
A.~Guo and M.~Haenggi, ``Spatial stochastic models and metrics for the
  structure of base stations in cellular networks,'' \emph{{IEEE} Trans.
  Wireless Commun.}, vol.~12, no.~11, pp. 5800--5812, Nov. 2013.

\bibitem{marco_fitting}
\BIBentryALTinterwordspacing
W.~Lu and M.~D. Renzo, ``Stochastic geometry modeling of cellular networks:
  Analysis, simulation and experimental validation,'' \emph{CoRR}, vol.
  abs/1506.03857, 2015. [Online]. Available:
  \url{http://arxiv.org/abs/1506.03857}
\BIBentrySTDinterwordspacing

\bibitem{survey_h}
H.~ElSawy, E.~Hossain, and M.~Haenggi, ``Stochastic geometry for modeling,
  analysis, and design of multi-tier and cognitive cellular wireless networks:
  A survey,'' \emph{{IEEE} Commun. Surveys Tuts.}, vol.~15, no.~3, pp.
  996--1019, 2013.

\bibitem{eid_Mimo}
M.~Di~Renzo and W.~Lu, ``Stochastic geometry modeling and performance
  evaluation of {MIMO} cellular networks using the equivalent-in-distribution
  {(EiD)}-based approach,'' \emph{{IEEE} Trans. Commun.}, vol.~63, no.~3, pp.
  977--996, Mar. 2015.

\bibitem{Laila_letter}
L.~H. Afify, H.~ElSawy, T.~Y. Al-Naffouri, and M.-S. Alouini, ``The influence
  of {Gaussian} signaling approximation on error performance in cellular
  networks,'' \emph{{IEEE} Commun. Lett.}, Accepted 2015.

\bibitem{Gil_marco}
M.~D. Renzo and P.~Guan, ``Stochastic geometry modeling of coverage and rate of
  cellular networks using the {Gil-Pelaez} inversion theorem.'' \emph{{IEEE}
  Commun. Lett.}, vol.~19, no.~9, pp. 1575--1578, Sep. 2014.

\bibitem{Rate_marco}
M.~D. Renzo, A.~Guidotti, and G.~E. Corazza, ``Average rate of downlink
  heterogeneous cellular networks over generalized fading channels: A
  stochastic geometry approach,'' \emph{{IEEE} Trans. Commun.}, vol.~61, no.~7,
  pp. 3050--3071, Jul. 2013.

\bibitem{Norman_MSC}
N.~Beaulieu and X.~Jiandong, ``A novel fading model for channels with multiple
  dominant specular components,'' \emph{{IEEE} Wireless Commun. Lett.}, vol.~4,
  no.~1, pp. 54--57, Feb 2015.

\bibitem{Yacoub}
M.~Yacoub, ``Nakagami-m phase-envelope joint distribution: A new model,''
  \emph{{IEEE} Trans. Veh. Technol.}, vol.~59, no.~3, pp. 1552--1557, March
  2010.

\bibitem{Rapaport}
G.~D. Durgin, T.~S. Rappaport, and D.~A. De~Wolf, ``New analytical models and
  probability density functions for fading in wireless communications,''
  \emph{{IEEE} Trans. Commun.}, vol.~50, no.~6, pp. 1005--1015, 2002.

\bibitem{GTR_just1}
J.~Frolik, ``On appropriate models for characterizing hyper-rayleigh fading,''
  \emph{{IEEE} Trans. Wireless Commun.}, vol.~7, no.~12, pp. 5202--5207, Dec.
  2008.

\bibitem{Slim_GTW}
M.~Rao, F.~Lopez-Martinez, M.-S. Alouini, and A.~Goldsmith, ``{MGF} approach to
  the analysis of generalized two-ray fading models,'' \emph{{IEEE} Trans.
  Wireless Commun.}, vol.~14, no.~5, pp. 2548--2561, May 2015.

\bibitem{Rappaport_mmW}
S.~Rangan, T.~Rappaport, and E.~Erkip, ``Millimeter-wave cellular wireless
  networks: Potentials and challenges,'' \emph{Proceedings of the IEEE}, vol.
  102, no.~3, pp. 366--385, March 2014.

\bibitem{Rapaport2}
T.~Rappaport, ``Characterization of {UHF} multipath radio channels in factory
  buildings,'' \emph{{IEEE} Trans. Antennas Propag.}, vol.~37, no.~8, pp.
  1058--1069, Aug 1989.

\bibitem{hamdi2010useful}
K.~A. Hamdi, ``A useful lemma for capacity analysis of fading interference
  channels,'' \emph{{IEEE} Trans. Commun.}, vol.~58, no.~2, pp. 411--416, 2010.

\bibitem{integrals_book}
I.~S. Gradshteyn and I.~M. Ryzhik, \emph{{Table of Integrals, Series, and
  Products, Seventh Edition}}.\hskip 1em plus 0.5em minus 0.4em\relax Academic
  Press, 2007.

\end{thebibliography}
\vfill
\end{document}